\documentstyle[12pt,epsfig,here,axodraw]{article}

\def\ti{\tilde}
\newcommand{\be}[1]{\begin{equation} \label{(#1)}}
\newcommand{\ee}{\end{equation}}
\newcommand{\baq}[1]{\begin{eqnarray} \label{(#1)}}
\newcommand{\eaq}{\end{eqnarray}}
\newcommand{\rf}[1]{(\ref{(#1)})}
\newcommand{\ba}{\begin{array}}
\newcommand{\ea}{\end{array}}
\newcommand{\T}{\tilde\tau}
\newcommand{\slashed}[1]{\not\!#1}

\newcommand{\CH}{\tilde\chi}
\def\lsim{\raise0.3ex\hbox{$\;<$\kern-0.75em\raise-1.1ex\hbox{$\sim\;$}}}
\def\gsim{\raise0.3ex\hbox{$\;>$\kern-0.75em\raise-1.1ex\hbox{$\sim\;$}}}

\begin{document}

\begin{flushright} UWThPh-2003-10
 \\WUE-ITP-2003-007 \\
   IFIC/03-30\\
   hep-ph/0306304
 \end{flushright} 
\begin{center}  
{\Large{\bf T-odd Correlations
		in the Decay of  Scalar Fermions 
}}\\[10mm]

{A.~Bartl$^1$, H.~Fraas$^2$, T.~Kernreiter$^{1,3}$,  
and O.~Kittel$^2$}
\vspace{0.3cm}\\ 

{\it$^1$Institut f\"ur Theoretische Physik, 
Univ.~Wien, A-1090 Vienna, \\ Austria \\}
{\it$^{2}$ Institut f\"ur Theoretische Physik und Astrophysik, 
Univ.~W\"urzburg,\\
D-97074, W\"urzburg, Germany \\}
{\it$^3$ Instituto de F\'\i sica Corpuscular--C.S.I.C., 
Univ. de \\Val\`encia, 
Val\`encia 46100, Spain}

\end{center}

\begin{abstract}

We define a CP sensitive asymmetry in the sfermion decays 
$\ti f \to f~\ti\chi^0_j~ \ell~\bar \ell~,
f~\ti\chi^0_j~ q~\bar q$, based on triple product
correlations between the momenta of the outgoing fermions. 
We study this asymmetry in the MSSM with complex parameters.
We show that the asymmetry is sensitive to the phases of the
parameters $\mu$ and $M_1$. 
The leading contribution stems from
the decay chain
$\ti f\to f~\ti\chi^0_j\to f~\ti\chi^0_1~Z\to 
f~\ti\chi^0_1~\ell~ \bar \ell~~ (f~\ti\chi^0_1~q ~\bar q)$, 
for which we obtain analytic
formulae for the amplitude squared. The asymmetry can go up to
$3 \%$ for $\ti f\to f~\ti\chi^0_1~ \ell~\bar \ell$, and up to
$20 \%$ for $\ti f\to f~\ti\chi^0_1 ~q~\bar q$. We also estimate the 
rates necessary to measure the asymmetry.

\end{abstract}

\section{Introduction}
In the standard model (SM) the source
of CP violation is given by the phase in the
Kobayashi-Maskawa matrix \cite{Kobayashi:fv}.
However, it has been argued that this source
is not enough to explain the observed baryon
asymmetry of the universe (see for example \cite{BGSM}) 
and new sources of CP violation have to be introduced.
In the minimal supersymmetric extension of the
SM (MSSM), several supersymmetric (SUSY) breaking 
parameters and the higgsino mass parameter can be complex.

The  phases of the SUSY parameters 
are restricted by the experimental upper limits 
on the electric dipole moments (EDMs) \cite{Masiero:xj}
of electron, neutron and the $^{199}$Hg and $^{205}$Tl atoms. 
The limiting bounds are
$|d_e| < 4.3\times10^{-27} ecm$ \cite{boundele},
$|d_n| < 6.3\times10^{-26} ecm$ \cite{boundneu},
$|d_{Hg}| < 2.1\times10^{-28} ecm$ \cite{boundHg} and
$|d_{Tl}| < 1.3\times10^{-24} ecm$ \cite{boundTl}, respectively.
The general consensus is that one of 
the following three conditions has to be realized: i) The phases 
are severely suppressed \cite{abeletal, smallphases}, 
ii) supersymmetric particles of the first two generations are 
rather heavy, with masses of the order of a TeV 
\cite{Dimopoulos:1995mi}, iii) there 
are strong cancellations between the different SUSY contributions 
to the EDMs \cite{cancel}. 
At one-loop level, for the electron EDM these are the neutralino and chargino 
contributions, and for the neutron EDM in addition also the 
gluino exchange contributes.
While the phase of $\mu$ is restricted, 
$|\varphi_{\mu}| \lsim \pi/10 $, 
there is no such restriction on the phase of $M_1$ \cite{Barger:2001nu}.
 
In order to clarify the situation an unambiguous 
determination of the SUSY phases is necessary.
In particular, for determining also the sign of the 
phases, measurements of CP sensitive 
observables are necessary.
The SUSY phases  give rise to
CP-odd (T-odd) observables already at tree level 
\cite{Choi:1999cc,Kizukuri:zp,Bartl:2002hi,Oshimo}.
An important class of such observables involves triple product 
correlations \cite{Valencia:1994zi,Gavela:1988jx}. 
They allow us to define various CP asymmetries which are
sensitive to the different CP phases.
These observables could be measured at future 
linear collider experiments \cite{TDR}
and would allow us to independently
determine the values of the phases.

In this paper we consider a T-odd correlation in the decays
\be{eq:decaychain}
\ti f \to f~\ti\chi^0_j~ \ell~\bar \ell~,
f~\ti\chi^0_j~ q~\bar q,
\ee
with $\ell=e,\mu,\tau$, and $q$ denotes a quark.
$ $
The T-odd correlation for the leptonic decay is
defined as
\be{eq:tripleprodlep}
O^{\ell}_{\rm odd}=
{\bf p}_{f}\cdot({\bf p}_{\ell}\times{\bf p}_{\bar \ell}),
\ee
and that for the hadronic decay as
\be{eq:tripleprodquark}
O^{q}_{\rm odd}=
{\bf p}_{f}\cdot({\bf p}_{q}\times{\bf p}_{\bar q}),
\ee
where ${\bf p}$ denotes the three-momentum of the corresponding
fermion.
We define the corresponding T-odd asymmetries as
\be{eq:Toddasym}
{\mathcal A}_{\rm T}^{\ell,q}=
\frac{\Gamma(O^{\ell,q}_{\rm odd}>0)-\Gamma(O^{\ell,q}_{\rm odd}<0)}
{\Gamma(O^{\ell,q}_{\rm odd}>0)+\Gamma(O^{\ell,q}_{\rm odd}<0)}=
\frac{\int {\rm Sign}[O^{\ell,q}_{\rm odd}]|
{\mathcal M}^{\ell,q}|^2{\rm dLips}}
	{\int |{\mathcal M}^{\ell,q}|^2{\rm dLips}},
\ee
where ${\mathcal M}^{\ell,q}$ is the matrix element
for the decay \rf{eq:decaychain}.
For the measurement of ${\mathcal A}_{\rm T}^{\ell}$
or ${\mathcal A}_{\rm T}^q$ it is necessary to
be able to distinguish between the charges
of ${\ell}^+$ and ${\ell}^-$ or $q$ and $\bar q$.
In the case $\ell=e,\mu,\tau$ this should be
possible experimentally on an event by event basis
at an $e^+e^-$ linear collider \cite{TDR}.
${\mathcal A}_{\rm T}^q$ may be measureable 
in the case of $q=c,b$, where flavour reconstruction 
is possible \cite{Damerell:1996sv}. However,
this will only be possible statistically for a given event sample.

\begin{figure}	
\scalebox{1.3}{
\begin{picture}(190,95)(-85,-20)
\ArrowLine(40,50)(0,50)
\put(115,73){$\tilde{\chi}_1^0$}
\Vertex(40,50){4}
\put(37,57){$\tilde f $}
\ArrowLine(40,50)(80,50)
\put(0,56){$ f $}
\ArrowLine(80,50)(110,75)
\put(58,58){$ \tilde{\chi}_j^0 $}
\DashLine(80,50)(105,50){1.5}
\ArrowArcn(80,50)(17,0,305)
\put(100,37){$ \theta_1$}
\Photon(80,50)(100,20){2}{5}
\put(75,30){$ Z $}
\Vertex(80,50){2}
\ArrowLine(125,15)(100,20)
\put(135,12){$\bar \ell~(\bar q)$}
\ArrowArc(100,20)(20,310,346)
\put(120,3){$ \theta_2$}
\ArrowLine(100,20)(85,0)
\put(60,-3){$ \ell ~(q) $}
\DashLine(100,20)(110,5){1.5}
\Vertex(100,20){2}
\end{picture}
}
\caption{Schematic picture of the subsequent two-body decays
	$\ti f \to f \ti\chi^0_j,~
\ti\chi^0_j \to Z\ti\chi^0_1,~
	Z \to \ell \bar\ell ~(q \bar q)$ 
in the $\ti f$ rest frame.\label{picstaudec}}
\end{figure}
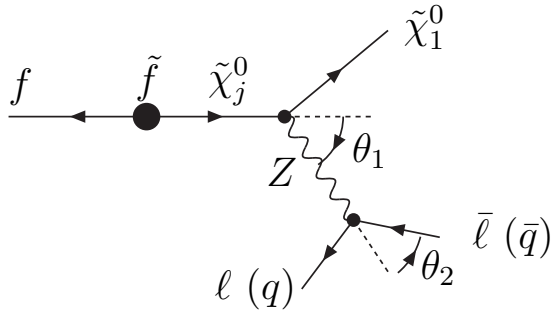

The leading contribution
to the triple products \rf{eq:tripleprodlep} and 
\rf{eq:tripleprodquark} 
originates from the decay chain
\be{eq:chidecay}
\ti f\to f\ti\chi^0_j\to f \ti\chi^0_1~Z\to 
f\ti\chi^0_1~\ell~\bar \ell 
~~(f\ti\chi^0_1~q~\bar q),
\ee
which is shown schematically in Fig.~\ref{picstaudec}.
Essentially, the triple products \rf{eq:tripleprodlep}
and \rf{eq:tripleprodquark}
are correlations between the $\ti\chi^0_j$ polarization 
and the $Z$ boson polarization, which are encoded in the momentum
vectors of the final leptons or quarks.
The correlations would vanish, for example, 
if a scalar particle in place of the $Z$ boson is exchanged.
Final state interactions may also contribute to 
${\mathcal A}_{\rm T}^{\ell,q}$, however, they arise
only at one-loop level. We expect that such contributions 
are smaller than 10\%, because only weak corrections 
to the absorptive part of the $\ti\chi^0_j$-$Z$-$\ti\chi^0_1$ 
vertex have to be included. 
In the similar case of the decay 
$\ti\chi^{\pm}_j\to W^{\pm}\ti\chi^0_1$,
corrections smaller than 10\% have been obtained \cite{Yang:2002am}.
Corrections of this order of magnitude have also been found in
\cite{Korner:2003zq}, where next-to-leading order effects 
on polarization observables within the SM have been studied.

As will be shown,
the tree-level contribution to the triple product
correlations \rf{eq:tripleprodlep}
and \rf{eq:tripleprodquark} are proportional to the imaginary
part of the $\ti\chi^0_j$-$Z$-$\ti\chi^0_1$ coupling squared
and are sensitive to the phases of the neutralino
mass parameters $M_1$ and $\mu$ (see Eqs. 
\rf{eq:matrixelement}-\rf{eq:matrixelement2}).
$O^{\ell,q}_{\rm odd}$ is not sensitive to the trilinear 
scalar coupling
parameter $A_f$ of the sfermion $\ti f$.
The reason is that the first decay in the chain
\rf{eq:chidecay}, $\ti f\to f\chi^0_j$, is a two body
decay of a scalar particle.
In order to be sensitive to the phase of $A_f$
one would have to construct instead of \rf{eq:tripleprodlep}, 
\rf{eq:tripleprodquark} a 
triple product correlation involving the
transverse polarization of the fermion $f$.
In principle this could be possible for
$f=\tau, t$ \cite{nojiri}, but this case will not be considered
here (for a similar case see \cite{Bartl:2002hi}).
The T-odd correlation \rf{eq:tripleprodlep} was 
proposed in \cite{Oshimo} and the size of the asymmetry was calculated
for the decay $\ti\mu\to\mu~\ti\chi^0_2\to
\chi^0_1~\mu~\ell^+~\ell^-$, however, for a
specific final state configuration only.
In the present paper we extend the work of  \cite{Oshimo}
by calculating the asymmetries \rf{eq:Toddasym} in the whole phase space.

In Section \ref{Definitions} we give the definitions and 
the matrix element of
the decay we are interested in.
In Section \ref{toddasymm} we perform 
the calculation of the T-odd asymmetry.
In Section \ref{numerics} we present our 
numerical results.
Section \ref{summary} contains a short summary and conclusion.

\section{Definitions and Formalism
  \label{Definitions} }  

\subsection{Lagrangian and Couplings}

The parts of the interaction Lagrangian of the MSSM
relevant for decay \rf{eq:chidecay} are (in our notation and conventions
we follow closely \cite{Haber:1984rc, thomas})
\be{eq:LagStauchi}
{\mathcal L}_{\ti f f\CH_j^0}=  \ti f_k \bar f
(b^{\ti f}_{kj} P_L+a^{\ti f}_{kj} P_R)\CH^0_j + {\rm h.c.}~,
\; j =1,\dots,4~, \; k=1,2~,
\ee
where
\be{eq:coupl1}
a_{kj}^{\ti f}=
g({\mathcal R}^{\ti f}_{kn})^{\ast}{\mathcal A}^f_{jn},\qquad 
b_{kj}^{\ti f}=
g({\mathcal R}^{\ti f}_{kn})^{\ast}
{\mathcal B}^f_{jn},\qquad
(n=L,R)
\ee
\begin{equation}
{\mathcal A}^{\ell,q}_j=\left(\begin{array}{ccc}
f^{\ell,q}_{Lj}\\[2mm]
h^{\ell,q}_{Rj} \end{array}\right),\qquad 
{\mathcal B}^{\ell,q}_j=\left(\begin{array}{ccc}
h^{\ell,q}_{Lj}\\[2mm]
f^{\ell,q}_{Rj} \end{array}\right),
\label{eq:coupl2}
\end{equation}
with
\baq{eq:coupl}
h^\ell_{Lj}&=& (h^\ell_{Rj})^{\ast}=Y_{\ell} N_{j3}^{\ast},\nonumber\\
f^\ell_{Lj}&=& -\frac{1}{\sqrt{2}}(\tan\Theta_W N_{j1}+N_{j2}),\nonumber\\
f^\ell_{Rj}&=& \sqrt{2}\tan\Theta_W N_{j1}^{\ast},
\eaq
\baq{eq:coupu}
h^u_{Lj}&=& (h^u_{Rj})^{\ast}=Y_u N_{j4}^{\ast},\nonumber\\
f^u_{Lj}&=& \frac{1}{\sqrt{2}}(\tan\Theta_W N_{j2}+N_{j1}),\nonumber\\
f^u_{Rj}&=& -\frac{2 \sqrt{2}}{3}\tan\Theta_W N_{j1}^{\ast},
\eaq
\baq{eq:coupd}
h^d_{Lj}&=& (h^d_{Rj})^{\ast}=Y_d N_{j3}^{\ast},\nonumber\\
f^d_{Lj}&=& -\frac{1}{\sqrt{2}}(\frac{1}{3}
\tan\Theta_W N_{j2}-N_{j2}),\nonumber\\
f^d_{Rj}&=& \frac{\sqrt{2}}{3}\tan\Theta_W N_{j1}^{\ast},
\eaq
and
\be{eq:yukawa}
Y_{\ell,d}= \frac{m_{\ell,d}}{\sqrt{2}m_W \cos\beta},\quad 
Y_u= \frac{m_u}{\sqrt{2}m_W \sin\beta}~.
\ee
Here, $P_{L,R}=1/2(1\mp\gamma_5)$,
$g$ denotes the weak coupling constant, $\Theta_W$ is
the weak mixing angle, 
$m_W$ is the mass of the $W$ boson
and ${\mathcal R}^{\ti f}_{kn}$ is the scalar fermion mixing 
matrix defined below.
$m_{\ell}$ and $m_{u}~(m_{d})$ is the mass of the corresponding
lepton and up-type (down-type) quark, respectively.  
$N_{ij}$ is the complex unitary $4\times 4$ matrix which diagonalizes
the neutral gaugino-higgsino mass matrix $Y_{\alpha\beta}$, 
$N_{i \alpha}^*Y_{\alpha\beta}N_{k\beta}^{\ast}=
m_{\tilde{\chi}^0_i}\delta_{ik}$,
in the basis ($\tilde{B},
\tilde{W}^3, \tilde{H}^0_1, \tilde{H}^0_2$) \cite{Haber:1984rc}.
The masses and couplings of the
$\ti f_k$ follow from the 
hermitian sfermion
mass matrix which in the basis 
$(\ti f_L, \ti f_R)$ reads
\begin{equation}
{\mathcal{L}}_M^{\ti f}= -(\ti f_L^{\dagger},\, \ti f_R^{\dagger})
\left(\begin{array}{ccc}
M_{\ti f_{LL}}^2 & e^{-i\varphi_{\ti f}}|M_{\ti f_{LR}}^2|\\[5mm]
e^{i\varphi_{\ti f}}|M_{\ti f_{LR}}^2| & M_{\ti f_{RR}}^2
\end{array}\right)\left(
\begin{array}{ccc}
\ti f_L\\[5mm]
\ti f_R \end{array}\right),
\label{eq:mm}
\end{equation}
with
\begin{eqnarray}
M_{\ti f_{LL}}^2 & = & M_{L\ti f}^2+(I^f_{3L}-q_f\sin^2\Theta_W)
\cos2\beta \ m_Z^2+m_f^2 , \label{eq:LL}\\[3mm]
M_{\ti f_{RR}}^2 & = & M_{R\ti f}^2+q_f\sin^2\Theta_W\cos2\beta \
m_Z^2+m_f^2 , \label{eq:RR} \\[3mm]
M_{\ti f_{RL}}^2 & = & (M_{\ti f_{LR}}^2)^{\ast}=
  m_f(A_f-\mu^{\ast}  
 (\cot\beta)^{2 I^f_{3L}}), \label{eq:mlr}
\end{eqnarray}
\begin{equation}
\varphi_{\ti f}  = \arg\lbrack A_f-\mu^{\ast}(\cot\beta)^{2 I^f_{3L}}\rbrack ,
\label{eq:phtau}
\end{equation}
where 
$m_Z$ is the mass of the $Z$ boson, $q_f$ and $I^f_{3 L,R}$ is the 
charge and the isospin of the 
fermion, respectively.
$M_{L\ti f}$, $M_{R\ti f}, A_f$ are the soft
SUSY--breaking parameters of 
the $\ti f_i$ system.
The $\ti f$ mass eigenstates are 
$(\tilde f_1, \tilde f_2)=(\ti f_L, \ti f_R)
{{\mathcal R}^{\ti f}}^{T}$ with 
 \begin{equation}
{\mathcal R}^{\T}
	 =\left( \begin{array}{ccc}
e^{i\varphi_{\ti f}}\cos\theta_{\ti f} & 
\sin\theta_{\ti f}\\[5mm]
-\sin\theta_{\ti f} & 
e^{-i\varphi_{\ti f}}\cos\theta_{\ti f}
\end{array}\right),
\label{eq:rtau}
\end{equation}
and
\begin{equation}
\cos\theta_{\ti f}=\frac{-|M_{\ti f_{LR}}^2|}{\sqrt{|M_{\ti f_{LR}}^2|^2+
(m_{\ti f_1}^2-M_{\ti f_{LL}}^2)^2}},\quad
\sin\theta_{\ti f}=\frac{M_{\ti f_{LL}}^2-m_{\ti f_1}^2}
{\sqrt{|M_{\ti f_{LR}}^2|^2+(m_{\ti f_1}^2-M_{\ti f_{LL}}^2)^2}}.
\label{eq:thtau}
\end{equation}
The mass eigenvalues are
\begin{equation}
 m_{\ti f_{1,2}}^2 = \frac{1}{2}\left((M_{\ti f_{LL}}^2+M_{\ti f_{RR}}^2)\mp 
\sqrt{(M_{\ti f_{LL}}^2 - M_{\ti f_{RR}}^2)^2 +4|M_{\ti f_{LR}}^2|^2}\right).
\label{eq:m12}
\end{equation}
The remaining parts of the interaction Lagrangian of the MSSM
relevant for the decay  \rf{eq:chidecay}
are
\be{eq:LagZchi}
{\mathcal L}_{Z\CH_i^0\CH_j^0}= Z^{\mu}
\overline{\CH^0_i}\gamma_{\mu}(O^{\prime\prime L}_{ij} P_L
+O^{\prime\prime R}_{ij} P_R) \CH^0_j~, \;\; i,j =1,\dots,4~,
\ee
and
\be{eq:LagZff}
{\mathcal L}_{Z f \bar f}=  Z^{\mu} \bar f
\gamma_{\mu}( L_f P_L+R_f P_R)f~,
\ee
respectively,
where
\baq{eq:Coupl}
O^{\prime\prime L}_{ij}&=&\frac{g}{4 \cos\Theta_W}(N_{i4} N_{j4}^{\ast} 
-N_{i3} N_{j3}^{\ast})~,\nonumber\\[3mm]
O^{\prime\prime R}_{ij}&=&-{O^{\prime\prime L}_{ij}}^{\ast}~.
\eaq
\be{eq:Zff}
L_f(R_f)=-\frac{g}{\cos\Theta_W}(I^f_{3 L(R)}- q_f \sin^2\Theta_W)~.
\ee
Note that
our definition of $O^{\prime\prime L,R}_{ij}$ ($L_f, R_f$)
differs from that given in \cite{Haber:1984rc} by a factor 
$g/2\cos\Theta_W$ $ (g/\cos\Theta_W)$.

\subsection{Spin-Density Matrix Formalism}

For the calculation of the amplitude squared of the 
subsequent two-body decays \rf{eq:chidecay}
of the sfermion, we use the spin-density matrix formalism
\cite{spinhaber, gudi}. The amplitude squared is given by
\be{eq:matrixelement}
|{\mathcal M}|^2=|\Delta(\ti\chi^0_j)|^2~|\Delta(Z)|^2
\sum_{\lambda_i,\lambda'_i,\lambda_k,\lambda'_k}~
{(\rho_{D_1})}_{\lambda_i\lambda'_i}~
{(\rho_{D_2})}^{\lambda'_i\lambda_i}_{\lambda_k\lambda'_k}~
{(\rho_{D_3})}^{\lambda'_k\lambda_k},
\ee
with the propagators  
$ \Delta(\ti\chi^0_j ) = 
1/[p^2_{\chi_j} -m^2_{\chi_j}
+im_{\chi_j}\Gamma_{\chi_j}]$ and
$ \Delta(Z) = 
1/[p^2_{Z} -m^2_{Z}
+im_{Z}\Gamma_{Z}]$.
Here, 
$p_{\chi_j}, m_{\chi_j}, \Gamma_{\chi_j}$
($p_{Z}, m_Z, \Gamma_Z$)
are the four-momenta, masses and widths of the decaying neutralino 
($Z$ boson), respectively.
The amplitude squared is composed of the unnormalized 
spin density matrices 
$\rho_{D_1}$, $\rho_{D_2}$ and $\rho_{D_3}$ of the
decay \rf{eq:chidecay}, which carry the
helicity indices $\lambda_i,\lambda'_i$ of the neutralinos 
and/or the helicity indices $\lambda_k,\lambda'_{k}$ of the $Z$ boson.
Introducing a set of polarization basis 4-vectors
$s^a_{\chi_j}\;(a=1,2,3)$ 
for the neutralino $ \ti\chi^0_j$, 
which fulfill the orthonormality relations 
$s^a_{\chi_j}\cdot s^b_{\chi_j}=-\delta^{ab}$ and
$s^a_{\chi_j}\cdot p_{\chi_j}=0$,
the density matrices can be 
expanded in terms of the Pauli matrices:
\baq{eq:rhoD} \label{eq:rhoD1}
{(\rho_{D_1})}_{\lambda_i\lambda'_i}&=&\delta_{\lambda_i\lambda'_i}~D_1+
\sigma^a_{\lambda_i\lambda'_i}~\Sigma^a_{D_1}~, \\
{(\rho_{D_2})}^{\lambda'_i,\lambda_i}_{\lambda_k,\lambda'_k}&=&
\left[\delta_{\lambda'_i\lambda_i}~D_2^{\mu\nu}+
\sigma^b_{\lambda'_i\lambda_i}~
\Sigma^{b\mu\nu}_{D_2}\right]
\varepsilon^{\lambda_k\ast}_{\mu}
\varepsilon^{\lambda'_k}_{\nu}~,\label{eq:rhoD2}\\
{(\rho_{D_3})}^{\lambda'_k\lambda_k}&=&
D_3^{\rho\sigma}~\varepsilon^{\lambda'_k\ast}_{\sigma}
\varepsilon_{\rho}^{\lambda_k}~.\label{eq:rhoD3}
\eaq
The polarization vectors 
$\varepsilon^{\lambda_k}_{\mu}$ of the $Z$ boson
obey $p^{\mu}_Z\varepsilon^{\lambda_k}_{\mu}=0$ and the completeness relation
$\sum_{\lambda_k} \varepsilon^{\lambda_k\ast}_{\mu}
\varepsilon^{\lambda_k}_{\nu}= -g_{\mu\nu}+p_{Z \mu}p_{Z \nu}/m_Z^2$.
The expansion coefficients of the density matrices 
(\ref{eq:rhoD1})-(\ref{eq:rhoD3}) are

\be{eq:D1}
D_1 = (|a^{\ti f}_{kj}|^2+|b^{\ti f}_{kj}|^2)(p_{f'} 
\cdot p_{\chi_j})~,
\ee
\be{eq:sigmaD1}
\Sigma^a_{D_1} =m_{\chi_j} (|b^{\ti f}_{kj}|^2-|a^{\ti f}_{kj}|^2)
(p_{f'}\cdot s^a_{\chi_j})~,
\ee
\baq{eq:D2}
{D_2}_{\rho\sigma}&=& 4 g_{\rho\sigma}\left[2~{\rm Re}
(O''^L_{1j} {O''^R_{1j}}^{\ast})
m_{\chi_1} m_{\chi_j} -(|O''^L_{1j}|^2+|O''^R_{1j}|^2)
(p_{\chi_1}\cdot p_{\chi_j})\right]
\\ [3mm]
&&{}+
4 (|O''^L_{1j}|^2+|O''^R_{1j}|^2)(p_{\chi_j\rho}p_{\chi_1\sigma}+
{p_{\chi_j}}_{\sigma}p_{\chi_1\rho})~,
\eaq
\baq{eq:S2}
\Sigma^{a}_{D_2\rho\sigma}&=& i~8~m_{\chi_1} 
{\rm Im}(O''^L_{1j}{O''^R_{1j}}^{\ast})(p_{\chi_j\rho}~
s^a_{\chi_j\sigma}-p_{\chi_j\sigma}~
s^a_{\chi_j\rho}){}\nonumber\\[3mm]
&&{}+i~4~\varepsilon_{\rho\sigma\mu\nu}~
p_{\chi_1}^{\mu}s^{a\nu}_{\chi_j}m_{\chi_j}(|O''^L_{1j}|^2+
|O''^R_{1j}|^2){}\nonumber\\[3mm]
&&{}-i~8~\varepsilon_{\rho\sigma\mu\nu}~
p_{\chi_j}^{\mu}s^{a\nu}_{\chi_j}m_{\chi_1}{\rm Re}
(O''^L_{1j}{O''^R_{1j}}^{\ast})~,
\eaq
\baq{eq:D3}
D_3^{\rho\sigma}&=&-2~g^{\rho\sigma}
(L_f^2+R_f^2)(p_{ f}\cdot p_{\bar f})
+2~(p^{\rho}_{f}~p^{\sigma}_{\bar f}+p^{\rho}_{\bar f}~p^{\sigma}_{f})
(L_f^2+R_f^2){}\nonumber\\[3mm]
{}&&+i~2~(R_f^2-L_f^2)~
\varepsilon^{\rho\sigma\mu\nu}~p_{f\mu}~p_{\bar f \nu}~,
\eaq
with $\varepsilon^{0123}=1$ and  $m_{\chi_1}$ the mass of the 
lightest supersymmetric particle (LSP).
The masses of the fermions $f=e,\mu,\tau,c,b$ are set to zero.
In Eqs.~\rf{eq:D1} and \rf{eq:sigmaD1} $f'$ denotes the fermion
steming from the first decay in Eq.~\rf{eq:chidecay}.
Inserting the density matrices (\ref{eq:rhoD1})-(\ref{eq:rhoD3})
in Eq. \rf{eq:matrixelement}, the amplitude squared is given by:
\be{eq:matrixelement2}
|{\mathcal M}|^2=2~|\Delta(\ti\chi^0_j)|^2~|\Delta(Z)|^2~
\lbrace D_1~D_{2 \rho\sigma}+
\Sigma^a_{D_1}~\Sigma^a_{D_2\rho\sigma}\rbrace D_3^{\rho\sigma}.
\ee

\section{T-odd Asymmetry
  \label{toddasymm}}

In the following we present in some detail the calculation of the
T-odd asymmetry in Eq.~\rf{eq:Toddasym} for the slepton decays
$\ti \ell\to \ell
\ti\chi^0_j\to \ell \ti\chi^0_1~Z\to \ell\ti\chi^0_1~f~\bar f$.
The replacements which must be made for the asymmetry
in the case of $\ti q$ decays
are obvious.

In the rest frame of $\ti \ell$, the coordinate system is defined 
such that the momentum four-vectors are given by
\baq{eq:momentumvec}
p_Z &=&(E_Z,0,0,|{\bf p}_Z|),\quad 
p_{\chi_j} =|{\bf p}_{\chi_j}|
(E_{\chi_j}/|{\bf p}_{\chi_j}|,\sin\theta_1,0,\cos\theta_1),\quad
{}\nonumber\\[3mm]
&&{}p_{\bar f} =
|{\bf p}_{\bar f}|(E_{\bar f}/|{\bf p}_{\bar f}|,\sin\theta_2\cos\phi_2,
\sin\theta_2\sin\phi_2,\cos\theta_2)~,
\eaq
where
\be{eq:threevec}
|{\bf p}_{\chi_j}|=\frac{m^2_{\ti \ell}-
m^2_{\chi_j}}{2~m_{\ti \ell}}~,\qquad
|{\bf p}_{\bar f}|=\frac{m^2_Z}{2(E_Z-|{\bf p}_Z|\cos\theta_2)}~,
\ee
and
\be{eq:threevec1}
|{\bf p}^{\pm}_Z|= \frac{
(m^2_{\chi_j}+m^2_Z-m^2_{\chi_1})|{\bf p}_{\chi_j}|
\cos\theta_1\pm E_{\chi_j}\sqrt{\lambda(m^2_{\chi_j},m^2_Z,m^2_{\chi_1})-
4|{\bf p}_{\chi_j}|^2~m^2_Z~(1-\cos^2\theta_1)}}
{2|{\bf p}_{\chi_j}|^2 (1-\cos^2\theta_1)+2 m^2_{\chi_j}}~,
\ee
with $\lambda(x,y,z)=x^2+y^2+z^2-2(x y+x z+y z)$.
There are two solutions $|{\bf p}^{\pm}_Z|$ in the case 
$|{\bf p}^0_{\chi_j}|<|{\bf p}_{\chi_j}|$,
where $|{\bf p}^0_{\chi_j}|= 
\sqrt{\lambda(m^2_{\chi_j},m^2_Z,m^2_{\chi_1})}/2m_Z$  
is the neutralino momentum if the $Z$ boson is produced at rest.
The $Z$ decay angle $\theta_1$ is constrained in that case 
and the maximal angle $\theta^{\rm max}_1$ is given by
\be{eq:maxangle}\sin\theta^{\rm max}_1=
\frac{|{\bf p}^0_{\chi_j}| }{|{\bf p}_{\chi_j}| }=
	\frac{m_{\ti \ell}}{m_Z}
\frac{\lambda^{\frac{1}{2}}(m^2_{\chi_j},m^2_Z,m^2_{\chi_1})}
{(m^2_{\ti \ell}-m^2_{\chi_j})}\leq 1~.
\ee
If $|{\bf p}^0_{\chi_j}|>|{\bf p}_{\chi_j}|$, the
decay angle $\theta_1$ is not constrained and there is
only the physical solution $|{\bf p}^+_Z|$ left.

The spin basis vectors of $\ti\chi^0_j$ 
in the $\ti \ell$ rest frame are chosen by
\be{eq:polvec}
s^1_{\chi_j}=\left(0,\frac{{\bf s}_2
\times{\bf s}_3}{|{\bf s}_2\times{\bf s}_3|}\right),\quad
s^2_{\chi_j}=\left(0,
\frac{{\bf p}_{\chi_j}\times{\bf p}_Z}{|{\bf p}_{\chi_j}
\times{\bf p}_Z|}\right),\quad
s^3_{\chi_j}=\frac{1}{m_{\chi_j}}
\left(|{\bf p}_{\chi_j}|, E_{\chi_j} \hat{\bf p}_{\chi_j} \right)~,
\ee
with $ \hat{\bf p}_{\chi_j} = {\bf p}_{\chi_j}/ |{\bf p}_{\chi_j}| $.
Together with $p^{\mu}_{\chi_j}/m_{\chi_j}$, the spin basis vectors
form an orthonormal set.

The Lorentz invariant phase space
element for the decay chain 
$\ti \ell\to \ell~\ti\chi^0_j\to \ell~\ti\chi^0_1~Z\to 
\ell~\ti\chi_1^0~{f}~{\bar f}$,
in the rest frame of $\ti \ell$, can be written as
\baq{eq:specificform}
{\rm dLips}(m^2_{\ti \ell},p_{\ell},p_{\chi_1},p_{\bar f},p_{f}) &=&
\frac{1}{(2\pi)^2}~{\rm dLips}(m^2_{\ti \ell},p_{\ell},p_{\chi_j})
{\rm d}s_{D_2} {}\nonumber\\[3mm]
{}&&
\times \sum_{\pm}
{\rm dLips}(s_{D_2},p_{\chi_1},p_Z^{\pm})
{\rm d}s_{D_3}{\rm dLips}(s_{D_3},p_{\bar f},p_{f})~,
\eaq
where $s_{D_2}=p^2_{\chi_j}$ and  
$s_{D_3}=p^2_Z$.
The Lorentz invariant phase space elements of the
sequence of two body decays read
\be{eq:dlips1}
{\rm dLips}(m^2_{\ti \ell},p_\ell,p_{\chi_j})=\frac{1}{8(2\pi)^2}
\left(1-\frac{m^2_{\chi_j}}{m^2_{\ti \ell}}\right)~{\rm d}\Omega~,
\ee
\be{eq:dlips2}
{\rm dLips}(s_{D_2},p_{\chi_1},p^{\pm}_Z)=\frac{1}{4(2\pi)^2}~
\frac{|{\bf p}^{\pm}_Z|^2}{|E_Z^{\pm}~|{\bf p}_{\chi_j}|\cos\theta_1-
E_{\chi_j}~|{\bf p}^{\pm}_Z||}~{\rm d}\Omega_1~,
\ee
\be{eq:dlips3}
{\rm dLips}(s_{D_3},p_{\bar f},p_{f})=\frac{1}{8(2\pi)^2}
\frac{m^2_Z}{(E_Z^{\pm}-|{\bf p}^{\pm}_Z|\cos\theta_2)^2}~{\rm d}\Omega_2~,
\ee
where ${\rm d}\Omega_i=\sin\theta_i~{\rm d}\theta_i~{\rm d}\phi_i$.

The partial $\ti \ell$ decay width for the decay chain
\rf{eq:chidecay} is given by
\be{eq:width}
\Gamma(\ti \ell\to \ell~\ti\chi^0_1~{f}~{\bar f})=
\frac{1}{2 m_{\ti \ell}}\int|{\mathcal M}|^2
{\rm dLips}(m^2_{\ti \ell},p_{\ell},p_{\chi_1},p_{\bar f},p_{f}).
\ee
We use the narrow width approximation for the propagators
$\Delta(\ti\chi^0_j)$ and $\Delta(Z):\\$
$\int|\Delta(\ti\chi^0_j)|^2 $ ${\rm d}s_{D_2} = 
\frac{\pi}{m_{\chi_j}\Gamma_{\chi_j}}, 
\int|\Delta(Z)|^2{\rm d}s_{D_3} = 
\frac{\pi}{m_{Z}\Gamma_{Z}}$.
The approximation for the neutralino propagator is 
justified for $(\frac{\Gamma_{\chi_j}}{m_{\chi_j}})^2\ll 1$,
which holds in our case with 
$\Gamma_{\chi_j}\lsim {\mathcal O}({\rm GeV}) $.

From Eqs.~\rf{eq:Toddasym} and \rf{eq:matrixelement2} we
obtain for the asymmetry
\be{eq:Adependence}
{\mathcal A}_{\rm T}^{\ell,q}=
\frac{\int |\Delta (\tilde\chi^0_j)|^2 |\Delta (Z)|^2
	{\rm Sign}[O^{\ell,q}_{\rm odd}]
	\Sigma^a_{D_1}~\Sigma^a_{D_2\rho\sigma}D_3^{\rho\sigma}
{\rm dLips}  }
{\int |\Delta (\tilde\chi^0_j)|^2 |\Delta (Z)|^2
	D_1~D_{2 \rho\sigma} D_3^{\rho\sigma}
{\rm dLips} }~,
\ee
where in the derivation of this expression we have used 
$ \int |\Delta (\tilde\chi^0_j)|^2 |\Delta (Z)|^2
{\rm Sign}[O^{\ell,q}_{\rm odd}] \\
D_1~D_{2 \rho\sigma} D_3^{\rho\sigma}
{\rm dLips}=0 $ in the numerator and 
$\int |\Delta (\tilde\chi^0_j)|^2 |\Delta (Z)|^2 $ $
\Sigma^a_{D_1}~\Sigma^a_{D_2\rho\sigma}D_3^{\rho\sigma}
{\rm dLips}=0$ in the denominator.
As can be seen from Eq.~\rf{eq:Adependence}, the asymmetry
${\mathcal A}_{\rm T}^{\ell,q}$ is proportional to the
spin correlation terms 
$\Sigma^a_{D_1}~\Sigma^a_{D_2\rho\sigma}D_3^{\rho\sigma}$.
In the spin correlations only the term which contains
the T-odd correlation $O^{\ell,q}_{\rm odd}$, Eqs.~ 
\rf{eq:tripleprodlep} and \rf{eq:tripleprodquark},
contributes to ${\mathcal A}_{\rm T}^{\ell,q}$.
The T-odd correlation $O^{\ell,q}_{\rm odd}$ is contained in 
the product of the first term of Eq.~\rf{eq:S2} and
the last term of Eq.~\rf{eq:D3}, which leads to
\be{eq:relpart}
\Sigma^{a}_{D_2\rho\sigma}D_3^{\rho\sigma}\supset
-32~m_{\chi_1} {\rm Im}(O''^L_{1j}{O''^R_{1j}}^{\ast})
(R_f^2-L_f^2)
\varepsilon^{\rho\sigma\mu\nu}~
p_{\chi_j\rho}~s^a_{\chi_j\sigma}~p_{f\mu}~p_{\bar f\nu}~.
\ee
In the rest frame of $\ti \ell$, 
$(p_\ell\cdot s^a_{\chi_j})=0$ for $a=1,2$,
hence, $\Sigma^{1,2}_{D_1}=0$ in Eq.~\rf{eq:sigmaD1} and 
only $\Sigma^{3}_{D_2\rho\sigma}$, defined in 
Eq.~\rf{eq:S2}, contributes to the spin correlation terms in the
total amplitude squared.
Using the explicit representation of the fermion
momentum vector, Eq.~\rf{eq:momentumvec}, and the neutralino spin
vector, Eq.~\rf{eq:polvec}, the term 
with the $\varepsilon$-tensor in \rf{eq:relpart} can be written as
\be{eq:epsilonpara}
\varepsilon^{\rho\sigma\mu\nu}~
p_{\chi_j\rho}~s^3_{\chi_j\sigma}~p_{f\mu}~p_{\bar f \nu}=
-m_{\chi_j}~{\bf \hat p}_\ell\cdot({\bf p}_{f}\times {\bf p}_{\bar f})=
-m_{\chi_j}~|{\bf p}_Z|~|{\bf p}_{\bar f}|\sin\theta_1~\sin\theta_2~
\sin\phi_2~,
\ee
where ${\bf \hat p}_\ell = {\bf p}_\ell/|{\bf p}_\ell|$.
Since $0\leq\theta_1,\theta_2\leq \pi$ the
sign of the correlation 
${\bf p}_\ell\cdot({\bf p}_{f}\times {\bf p}_{\bar f})$ is given
by the sign of $\sin\phi_2$.
Thus $O_{\rm odd}>0$ corresponds to an integration
$\int^{\pi}_{0} {\rm d}\phi_2$, while $O_{\rm odd}<0$
corresponds to an integration
$\int^{2\pi}_{\pi} {\rm d}\phi_2$.
We therefore integrate in 
Eq.~\rf{eq:width} over the entire phase space
except over the angle $\phi_2$.
The T-odd asymmetry can then be written as
\be{eq:toddasymangle}
{\mathcal A}_{\rm T}^f=
\frac{\left[\int^{\pi}_0\frac{{\rm d}\Gamma}{{\rm d}\phi_2}
		-\int^{2\pi}_{\pi}\frac{{\rm d}\Gamma}{{\rm d}\phi_2}\right]
{\rm d}\phi_2}
{\left[\int^{\pi}_0\frac{{\rm d}\Gamma}{{\rm d}\phi_2}
		+\int^{2\pi}_{\pi}\frac{{\rm d}\Gamma}{{\rm d}\phi_2}\right]
	{\rm d}\phi_2}~.
\ee
The dependence of ${\mathcal A}_{\rm T}^f$ 
on the $\ti \ell_k$-$\ell$-$\ti\chi^0_j$ couplings  
$a^{\ti \ell}_{kj},b^{\ti \ell}_{kj}$, 
on the $Z$-$\bar f$-$f$ couplings $L_f,R_f$ and on the 
$Z$-$\ti \chi_1^0$-$\ti \chi_j^0$ couplings $O''^{L,R}_{1j}$ is given by
\be{eq:prop1}
{\mathcal A}^f_{\rm T}\propto 
\frac{|a^{\ti \ell}_{kj}|^2-|b^{\ti \ell}_{kj}|^2}
{|a^{\ti \ell}_{kj}|^2+|b^{\ti \ell}_{kj}|^2}~
\frac{L_f^2-R_f^2}{L_f^2+R_f^2}~
{\rm Im}(O''^L_{1j}{O''^R_{1j}}^{\ast})~,
\ee
which follows  from  Eqs.~\rf{eq:Adependence} and \rf{eq:relpart}.
Due to the first factor
$\frac{|a^{\ti \ell}_{kj}|^2-|b^{\ti \ell}_{kj}|^2}
{|a^{\ti \ell}_{kj}|^2+|b^{\ti \ell}_{kj}|^2}$,
the asymmetry ${\mathcal A}^f_{\rm T}$ will be
strongly suppressed for the case 
$|a^{\ti\ell}_{kj}| \approx |b^{\ti\ell}_{kj}|$
and maximally enhanced in the case of vanishing mixing 
in the slepton sector
$\frac{|a^{\ti \ell}_{kj}|^2-|b^{\ti \ell}_{kj}|^2}
{|a^{\ti \ell}_{kj}|^2+|b^{\ti \ell}_{kj}|^2}\approx \pm1 $.
Due to the second factor
$\frac{L_f^2-R_f^2}{L_f^2+R_f^2}$,
${\mathcal A}^{b(c)}_{\rm T}$ is larger than 
${\mathcal A}^\ell_{\rm T}$, with 
\be{eq:prop2}
{\mathcal A}^{b(c)}_{\rm T}=
\frac{L_{\ell}^2+
R_{\ell}^2}
{L_{\ell}^2-R_{\ell}^2}
\frac{L_{b(c)}^2-
R_{b(c)}^2}
{L_{b(c)}^2+R_{b(c)}^2}~
{\mathcal A}^{\ell}_{\rm T}\simeq 6.3~(4.5)
\times{\mathcal A}^{\ell}_{\rm T}.\nonumber
\ee

Note that the r.h.s of \rf{eq:relpart} and therefore the asymmetry
in Eq.~\rf{eq:Toddasym}
vanish for $m_{\chi_1}\to 0$, which is related to the
fact that it is possible to redefine the Weyl spinor
$\chi_1\to e^{i\alpha}\chi_1$
in this limit.

\section{Numerical Results
         \label{numerics}}

We present numerical results for the T-odd asymmetry
${\mathcal A}^{\ell}_{\rm T}$ defined in Eq.~\rf{eq:Toddasym}.
The values for ${\mathcal A}^{b,c}_{\rm T}$ may be obtained
from Eq.~\rf{eq:prop2}.
We analyze numerically the  decay chain 
$\ti \tau\to \tau\ti\chi^0_j\to \tau \ti\chi^0_1~Z\to 
\tau\ti\chi^0_1~\ell~\bar \ell$, $\ell=e,\mu,\tau$
where $\ti\chi^0_1$ is the lightest supersymmetric particle (LSP).
We assume that $\ti\tau_1$ is the lightest sfermion
and that the decays into a real $\ti\chi^0_j, \;j=2,3$,
and a real $Z$ are kinematically possible.
In the numerical study below we will treat the two cases
$\ti\tau_1\to\tau\ti \chi^0_2 $ and
$\ti\tau_1\to\tau\ti \chi^0_3 $ seperately.
The asymmetry ${\mathcal A}^{\ell,q}_{\rm T}$ in
Eq.~\rf{eq:Toddasym} could in principle also be
studied in $ \ti\chi^0_j$ three-body decays if the
two-body decays are kinematically forbidden \cite{Choi:1999cc},
which will be treated elsewhere \cite{karl}.

The relevant MSSM parameters are $
|\mu|, \varphi_{\mu},
|M_1|, \varphi_{M_1},
M_2, \tan\beta,
|A_{\tau}|, \varphi_{A_{\tau}}, m_{\ti\tau_1}, m_{\ti\tau_2}$ and the
Higgs mass parameter $m_{A}$.
For all scenarios we keep
$\tan\beta =10$,
$|A_{\tau}|=1000$~GeV, $\varphi_{A_{\tau}}=0$, 
$m_{\ti\tau_1}=300$~GeV, $ m_{\ti\tau_2}=800$~GeV
and use the GUT relation 
$|M_1|=5/3 \tan^2\Theta_W M_2$
in order to reduce the number of free parameters.
We have checked that our results do not depend
sensitively on this choice.
We choose  
$m_{A}=800$~GeV to rule out decays of  
the neutralino into charginos and
the charged Higgs bosons 
$\ti\chi^0_j \slashed{\to} \ti\chi^{\pm}_i H^{\mp},~i=1,2$,
as well as decays into the heavy neutral higgs
bosons $\ti\chi^0_j \slashed{\to} \ti\chi^0_i~H_{2,3}^0$.
For the calculation of the branching ratios
${\rm BR}(\ti \tau_1 \to \tau~\ti\chi^0_j)$ and
${\rm BR}(\ti\chi^0_j\to Z\ti\chi^0_1)$,
we take into account also the decays
$\ti\tau_1 \to \ti\chi^-_j\nu_{\tau},
\ti\chi^0_j\to H_1^0 \ti\chi^0_1,W^{\pm}\ti\chi^{\mp}_1$
in addition to 
$\ti\tau_1 \to  \ti\chi^0_j\tau, \ti\chi^0_j \to Z \ti\chi^0_1$.
$H_1^0$ is the lightest neutral Higgs boson, which
in general is not a CP eigenstate
\cite{Pilaftsis:1998dd,carenaetal,dreescp}.
The decay $\ti\tau_1\to\tau\ell~\ti \ell$, $\ell=e,\mu$
is kinematically forbidden due to our assumption that
$\ti\tau_1$ is the lightest sfermion.
Other decay chains leading to the same final state are 
less important and will be neglected.

\subsection{Decay chain via $\ti\chi^0_2$}

\begin{figure}[H]
\begin{picture}(120,220)(0,0)
\put(10,0){\includegraphics{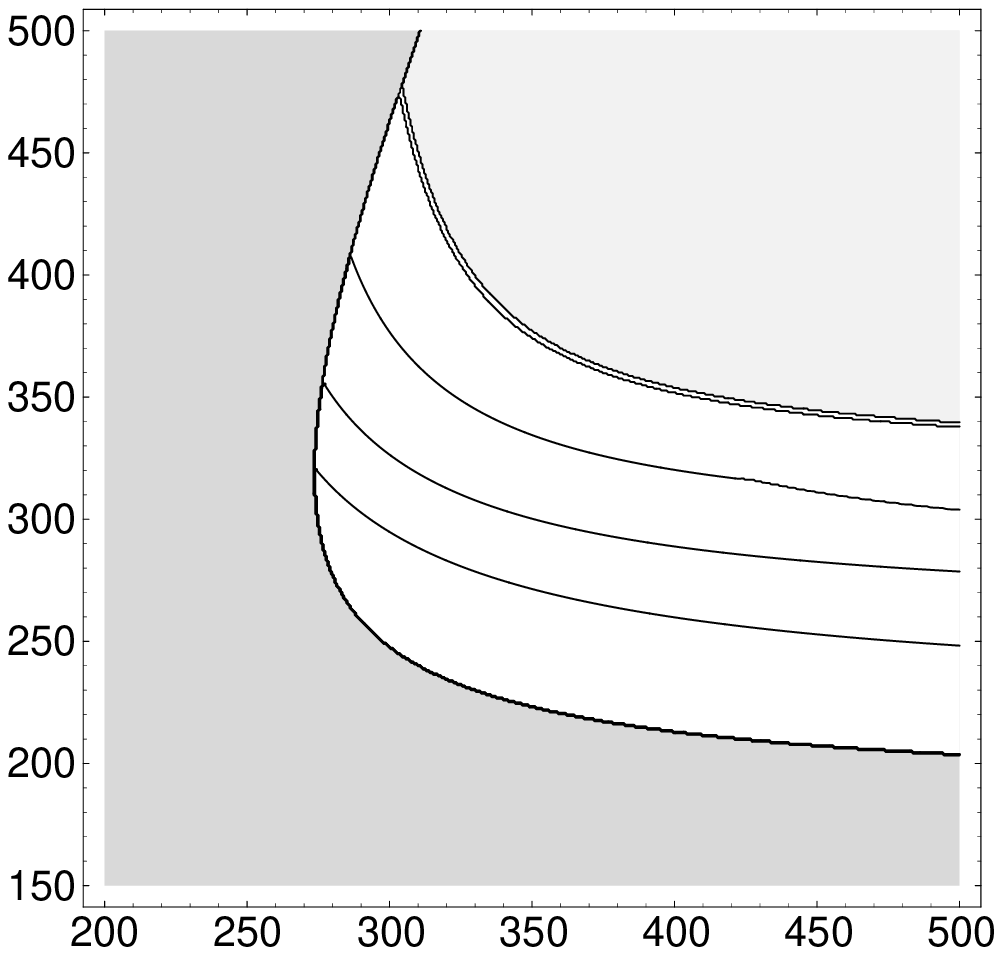}}
   \put(175,-8){$\mu$/GeV}
	\put(10,210){$M_2 /GeV$}
	\put(70,210){\fbox{BR($\tilde{\tau}_1 \;\to \;\tau \;
			\tilde{\chi}^0_1\;\ell \;\bar \ell
		$) in \% }}
	\put(110,117){{\footnotesize $1$}}
	\put(135,93){{\footnotesize $2 $}}
	\put(160,75){{\footnotesize $2.5 $}}
	\put(30,-8){Fig.~\ref{stau2a}a}
\put(240,0){\includegraphics{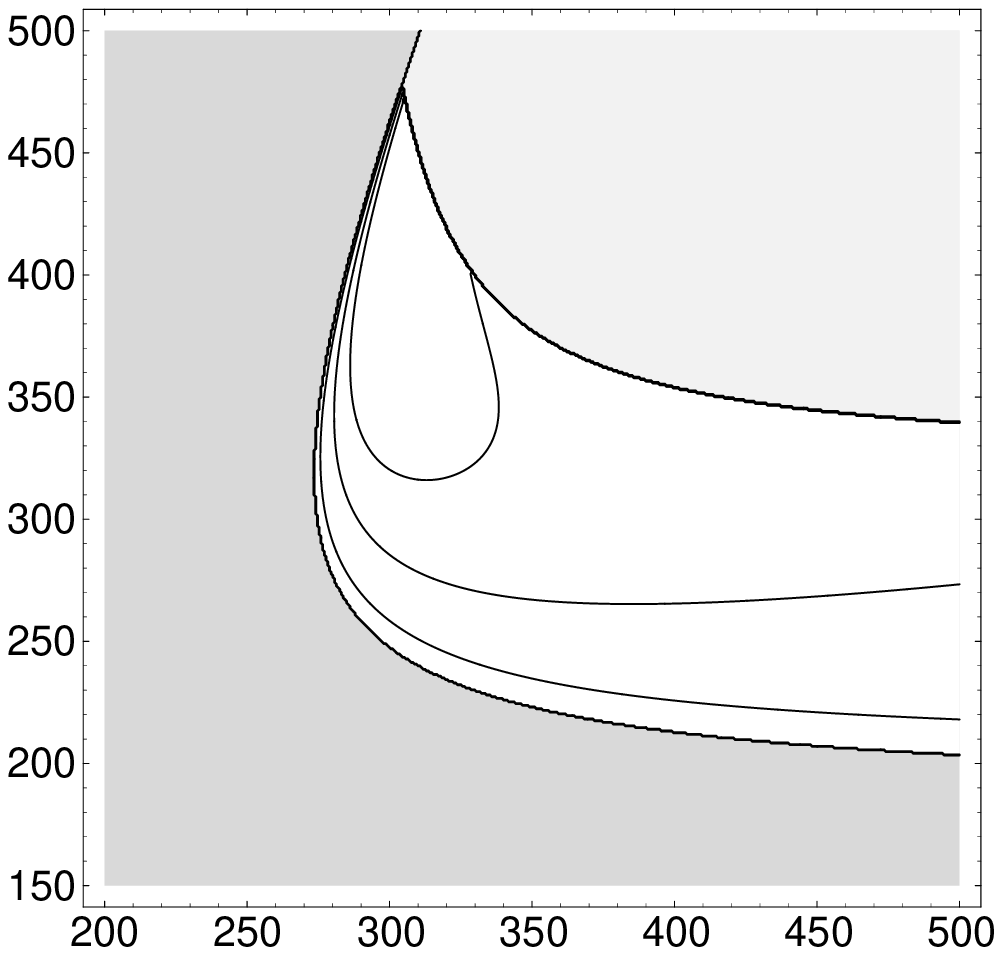}}
   \put(405,-8){$\mu$/GeV}
	\put(240,210){$M_2 /GeV$}
	\put(320,210){\fbox{${\mathcal A}_T^\ell$ in \% }}
   \put(323,106){{\footnotesize$3 $}}
   \put(350,80){{\footnotesize$2.5 $}}
	\put(380,60){{\footnotesize$1 $}}
	\put(260,-8){Fig.~\ref{stau2a}b }
 \end{picture}
 \caption{Contour lines of the branching ratio 
for $\ti\tau_1\to \ti\chi^0_1\tau \ell\bar\ell$ 
and asymmetry $A_T^\ell$ defined in Eq.~\rf{eq:Toddasym} 
in the $M_2$-$\mu$ plane for
	 $\varphi_{M_1}=\pi/2$ and $ \varphi_{\mu}=0$, taking
$\tan\beta =10$, $A_{\tau}=1000$~GeV, 
$m_{\ti\tau_1}=300$~GeV, $ m_{\ti\tau_2}=800$~GeV
for $M_{\ti E} > M_{\ti L}$. The grey areas
are kinematically forbidden since here 
$m_{\ti\tau_1}< m_{\chi^0_2}+m_{\tau}$ (light grey) or 
$m_{\chi^0_2}<m_{\chi^0_1}+m_Z$ (dark grey).
\label{stau2a}}
\end{figure}

First we consider ${\mathcal A}^\ell_{\rm T}$ for the decay chain
$\ti \tau_1 \to \tau~\ti\chi^0_2 
\to \tau Z~\ti\chi^0_1 \to \tau \ti\chi^0_1\ell~\bar \ell$,
for $\ell=e,\mu,\tau$.
In Fig.~\ref{stau2a}a we show the contour
lines for the branching ratio BR$(\tau_1 \to \tau~\ti\chi^0_1~\ell~\bar \ell) 
={\rm BR}(\ti \tau_1 \to \tau~\ti\chi^0_2)\times
{\rm BR}(\ti\chi^0_2\to Z\ti\chi^0_1)\times
{\rm BR}(Z \to \ell~\bar \ell)$
in the $M_2$-$|\mu|$ plane
for $\varphi_{M_1}=\pi/2$ and $ \varphi_{\mu}=0$.
Small values of the phase of $\mu$
are suggested by constraints on the EDMs 
for a typical SUSY scale of the order of a few 100 GeV
\cite{Barger:2001nu}.
For BR$(\tau_1 \to \tau~\ti\chi^0_1~\ell~\bar \ell)$ we always sum over
the lepton anti-lepton pairs which couple to the 
$Z$.
The grey areas in  Fig.~\ref{stau2a}
are kinematically forbidden since here 
$m_{\ti\tau_1}< m_{\chi^0_2}+m_{\tau}$ (light grey) or 
$m_{\chi^0_2}<m_{\chi^0_1}+m_Z$ (dark grey).	
We choose $M_{\ti E} > M_{\ti L}$ since in this case the 
$\ti \tau_1$-$\tau$-$\ti\chi^0_2$ coupling $|a^{\ti\tau}_{12}|$ is larger, 
which implies a larger branching ratio 
BR$(\ti \tau_1 \to \tau~\ti\chi^0_2)$
than for $M_{\ti E} < M_{\ti L}$. 
(We use the usual notation 
$M_{\ti E}\equiv M_{R\ti\tau}$, $M_{\ti L}\equiv M_{L\ti\tau}$,
see Eqs.~(\ref{eq:LL}) and (\ref{eq:RR})).
$M_{\ti E} > M_{\ti L}$ is suggested in some scenarios
with non-universal scalar mass parameters at the GUT scale  
\cite{Baer:2000cb}. Furthermore, in Eqs.~(\ref{eq:LL}) and (\ref{eq:RR}) 
one could have $M_{\ti\tau_{LL}}< M_{\ti\tau_{RR}}$ 
in extended models with additional D-terms \cite{Hesselbach:2001ri}.
In a large region of the parameter
space we have BR$(\ti\chi^0_2\to Z\ti\chi^0_1) = 1$,
and we take BR$(Z \to \ell~\bar \ell)=0.1$.
The corresponding asymmetry ${\mathcal A}_{\rm T}^\ell$
is shown in Fig.~\ref{stau2a}b.
The dependence of ${\mathcal A}_{\rm T}^\ell$ on $M_2$ and $|\mu|$ 
is dominantly determined by 
${\rm Im}(O''^L_{12}{O''^R_{12}}^{\ast})$.

\begin{figure}[H]
\begin{picture}(120,220)(0,0)
\put(10,0){\includegraphics{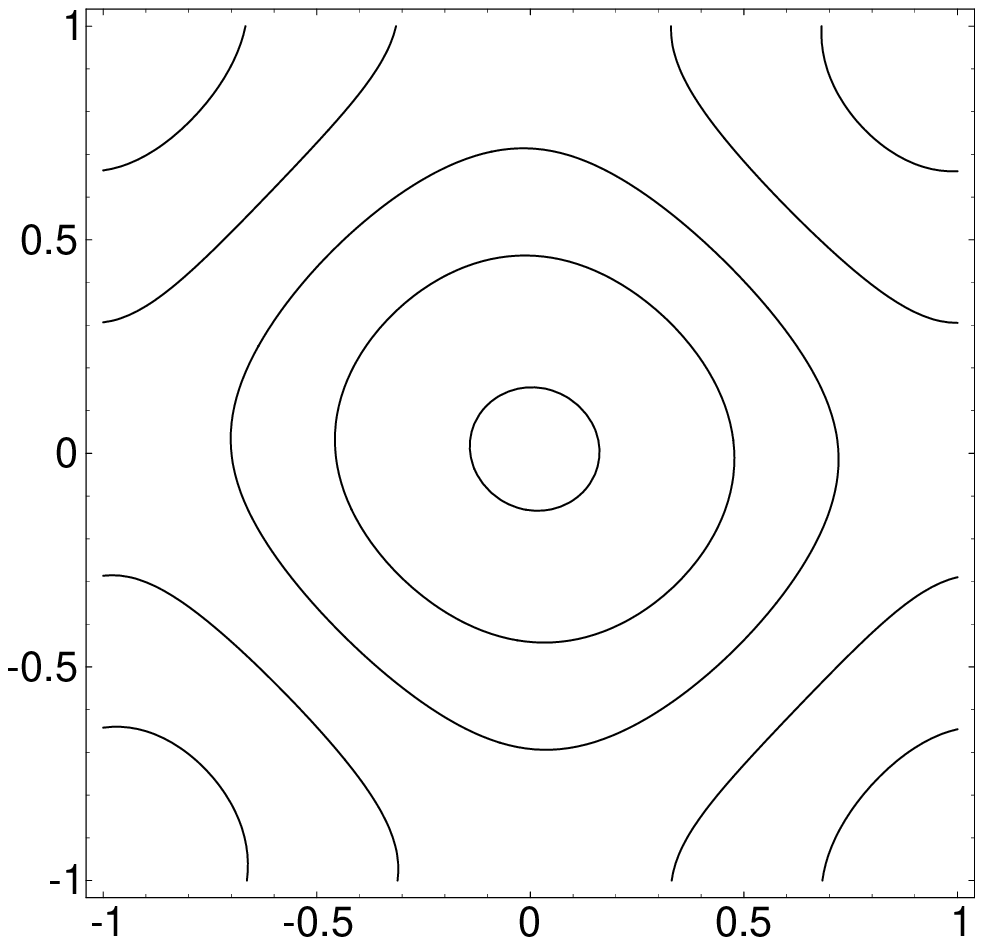}}
   \put(175,-8){$\varphi_{\mu}/\pi $}
	\put(10,210){$\varphi_{ M_1}/\pi$}
	\put(70,210){\fbox{BR($\tilde{\tau}_1 \;\to \;\tau  \;
			\tilde{\chi}^0_1\;\ell \; \bar \ell
		$) in \% }}
	\put(118,102){{\footnotesize $2.8 $}}
	\put(139,81){{\footnotesize $2.6 $}}
	\put(153,70){{\footnotesize$2.4 $}}
	\put(167,59){{\footnotesize$2.2 $}}
	\put(176,37){{\footnotesize$2.0 $}}
        \put(120,45){$\otimes$}
        \put(162,148){$\bullet$}
	\put(30,-8){Fig.~\ref{stau2b}a}
\put(240,0){\includegraphics{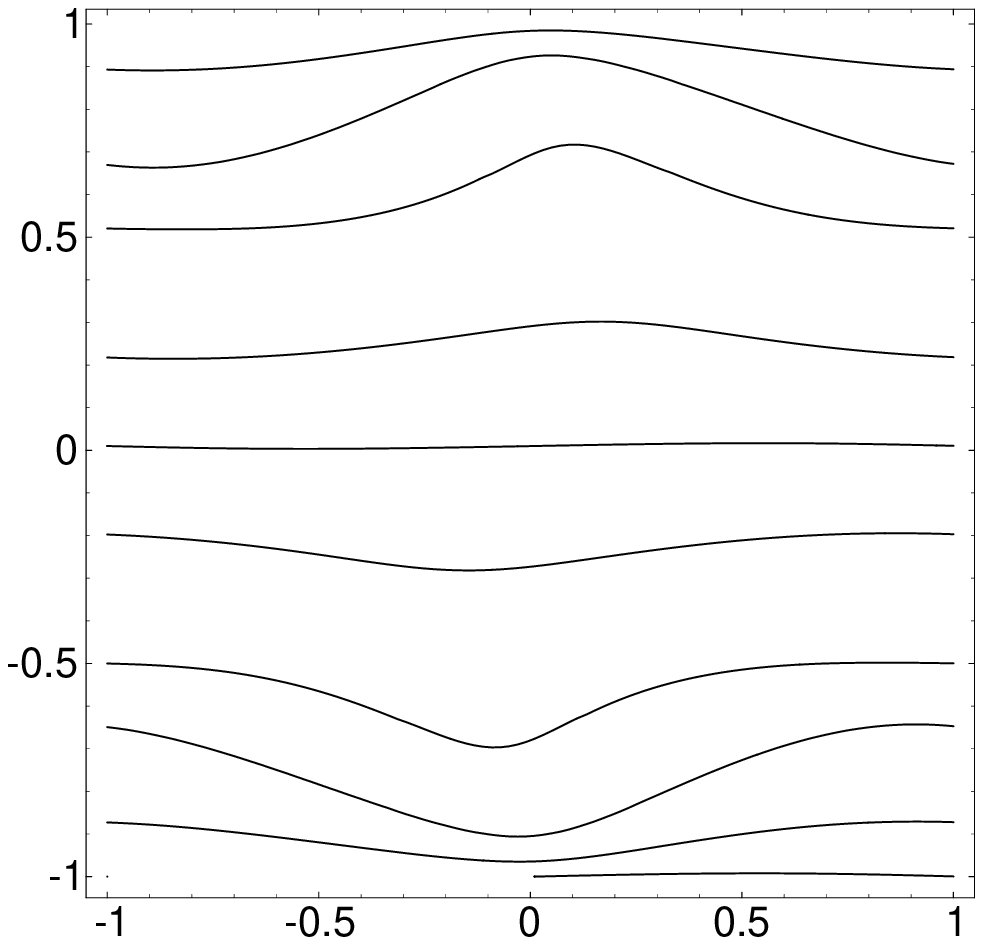}}
   \put(405,-8){$\varphi_{\mu}/\pi $}
	\put(240,210){$\varphi_{ M_1}/\pi$}
	\put(320,210){\fbox{${\mathcal A}_T^\ell$ in \% }}
   \put(270,186){{\footnotesize $1.5 $}}
   \put(280,166){{\footnotesize $3 $}}
	\put(293,154){{\footnotesize $3 $}}
	\put(300,130){{\footnotesize $1.5 $}}
   \put(330,110){{\footnotesize $0 $}}
	\put(340,87){{\footnotesize $-1.5 $}}
	\put(353,56){{\footnotesize $-3 $}}
   \put(360,37){{\footnotesize $-3 $}}
	\put(395,33){{\footnotesize $-1.5 $}}
        \put(347,45){$\otimes$}
        \put(394,148){$\bullet$}
	\put(260,-8){Fig.~\ref{stau2b}b }
 \end{picture}
\caption{Contour lines of the branching ratio 
for $\ti\tau_1\to \ti\chi^0_1\tau \ell\bar\ell$ 
and asymmetry $A_T^\ell$ defined in Eq.~\rf{eq:Toddasym} 
in the $\varphi_{ M_1}$-$\varphi_{\mu} $ plane 
for $|\mu|=300$~GeV and $M_2=280$~GeV, taking
$\tan\beta =10$, $A_{\tau}=1000$~GeV, 
$m_{\ti\tau_1}=300$~GeV, $ m_{\ti\tau_2}=800$~GeV
for $M_{\ti E} > M_{\ti L}$.
The points denoted by $\bullet$ and $\otimes$, respectively, 
are for the theoretical estimate 
of the necessary number of produced $\T_1$'s (see text).
\label{stau2b}}
\end{figure}

In Fig.~\ref{stau2b} we show 
the $\varphi_{M_1}$ and $\varphi_{\mu}$ dependence 
of BR$(\ti\tau_1 \to \tau~\chi^0_1~\ell~\bar \ell)$
and of ${\mathcal A}_{\rm T}^\ell$ 
in the full range of the phases for
$|\mu|=300$~GeV and $M_2=280$~GeV.
We display in Table \ref{tab:chi2} the masses of $\ti\chi^0_i, i=1,\dots,4$ 
and the total widths $\Gamma_{\CH_2}$, $\Gamma_{\T_1}$ 
for various values of the phases.
The value of ${\mathcal A}_{\rm T}^\ell$ depends stronger on $\varphi_{M_1}$,
which also determines the sign of ${\mathcal A}_{\rm T}^\ell$,
than on $\varphi_{\mu}$.

Based on our results on the asymmetry 
${\mathcal A}^{\ell }_{\rm T}$ in $\ti\tau_1\to\tau\ti\chi^0_2\to
\ti\chi^0_1\tau\ell^+ \ell^-$ and the branching ratio
we give a theoretical estimate of the number of produced
$\ti\tau_1$'s necessary to observe the T-odd asymmetry
in Eq.~\rf{eq:Toddasym}.
The relevant quantity to decide whether ${\mathcal A}_{\rm T}^\ell$
is observable (at 1$\sigma$), is given by 
$(({\mathcal A}^{\ell }_{\rm T})^2\times {\rm BR})^{-1}$, where 
${\rm BR}$ stands for the branching ratio of the decay
considered.
The necessary number of produced $\ti\tau_1$'s should then be 
$\gsim (({\mathcal A}^{\ell }_{\rm T})^2\times {\rm BR})^{-1}$.
As an example we take the point denoted by $\bullet$ in Fig.~\ref{stau2b},
with $\varphi_{\mu}=\pi/2$ and $\varphi_{M_1} = \pi/2$.
For this point ${\rm BR} \approx 2.5\times 10^{-2}$
and $|{\mathcal A}_{\rm T}^\ell|\approx 3\times 10^{-2}$ which
implies that $(({\mathcal A}^{\ell}_{\rm T})^2\times {\rm BR})^{-1}
\approx 4.4\times10^{5}$.
For the decay $\ti \tau_1\to b\bar b\ti\chi^0_1\tau$, on the other
hand, ${\rm BR} \approx 3.6\times 10^{-2}$ and 
$|{\mathcal A}_{\rm T}^b|\approx 1.9 \times 10^{-1}$, 
so that $(({\mathcal A}^{b}_{\rm T})^2\times {\rm BR})^{-1}
\approx 7.7\times 10^{2}$.
For comparison we consider a further example with smaller CP violating phases
$\varphi_{\mu}=0$ and $\varphi_{M_1}= -0.3 \pi$
(denoted by $\otimes$ in Fig.~\ref{stau2b}).
Also in this case we obtain almost the same results for
$(({\mathcal A}^{\ell,b}_{\rm T})^2\times {\rm BR})^{-1}$.
In these two examples,
the asymmetry ${\mathcal A}_{\rm T}^{\ell,q}$
should be measurable at an $e^+ e^-$ linear
collider  with $\sqrt{s}=800$~GeV and an integrated
luminosity of $500~fb^{-1}$
for $m_{\ti\tau_1}=300$~GeV.
It is clear that detailed Monte Carlo studies taking into
account background and detector simulations are necessary to get a more 
precise prediction of the expected accuracy.
However, this is beyond the scope of the present paper.
For a Monte Carlo study on a T-odd observable in neutralino
production and decay see \cite{Choi:2003pq}.

\begin{table}[H]
	\caption{
Masses of $\ti\chi^0_i, i=1,\dots,4$ 
and widths $\Gamma_{\CH_2}$, $\Gamma_{\T_1}$
for various phase combinations 
of $\varphi_{\mu}$ and $\varphi_{M_1}$, taking $|\mu|=300$~GeV and 
$M_2=280$~GeV,  
$\tan\beta =10$, $A_{\tau}=1000$~GeV, 
$m_{\ti\tau_1}=300$~GeV, $ m_{\ti\tau_2}=800$~GeV for $M_{\ti E} > M_{\ti L}$.
}
 \label{tab:chi2}
\begin{center}
\begin{tabular}{|c|c|c|c|c|} \hline
$\varphi_{\mu}$ & $\varphi_{M_1}$ &
$m_{\CH_1},m_{\CH_2},m_{\CH_3},m_{\CH_4}  ~[\rm GeV]$  &
$\Gamma_{\CH_2}~[\rm MeV]$ & $\Gamma_{\T_1}~[\rm MeV]$\\ \hline\hline
0 &            0    &$135,\;\;234,\;\;306,\;\;358$& 4.06 & 527 \\
0 & $\frac{\pi}{2}$ &$137,\;\;233,\;\;308,\;\;357$& 1.79 & 550 \\
0 & $ \pi   $       &$138,\;\;231,\;\;309,\;\;356$& 0.09 & 573 \\ \hline
$\frac{\pi}{2}$ &           0     & $137,\;\;239,\;\;307,\;\;353$& 5.43 &
487  \\
$\frac{\pi}{2}$ & $\frac{\pi}{2}$ & $138,\;\;238,\;\;309,\;\;352$& 2.89 &
511  \\
$\frac{\pi}{2}$ & $\pi$           & $137,\;\;237,\;\;311,\;\;351$& 1.49  &
529  \\\hline
$\pi$ &       0         & $138,\;\;245,\;\;309,\;\;347$&7.25  & 448 \\
$\pi$ & $\frac{\pi}{2}$ & $137,\;\;244,\;\;311,\;\;346$&5.78  & 466 \\
$\pi$ & $\pi$           & $136,\;\;243,\;\;313,\;\;345$&4.32  & 484 \\
\hline
 \end{tabular}
\end{center}
\end{table}

\vspace{2cm}

\subsection{Decay chain via $\ti\chi^0_3$
\label{decaychain3}}

\begin{figure}[H]
\begin{picture}(120,220)(0,0)
\put(10,0){\includegraphics{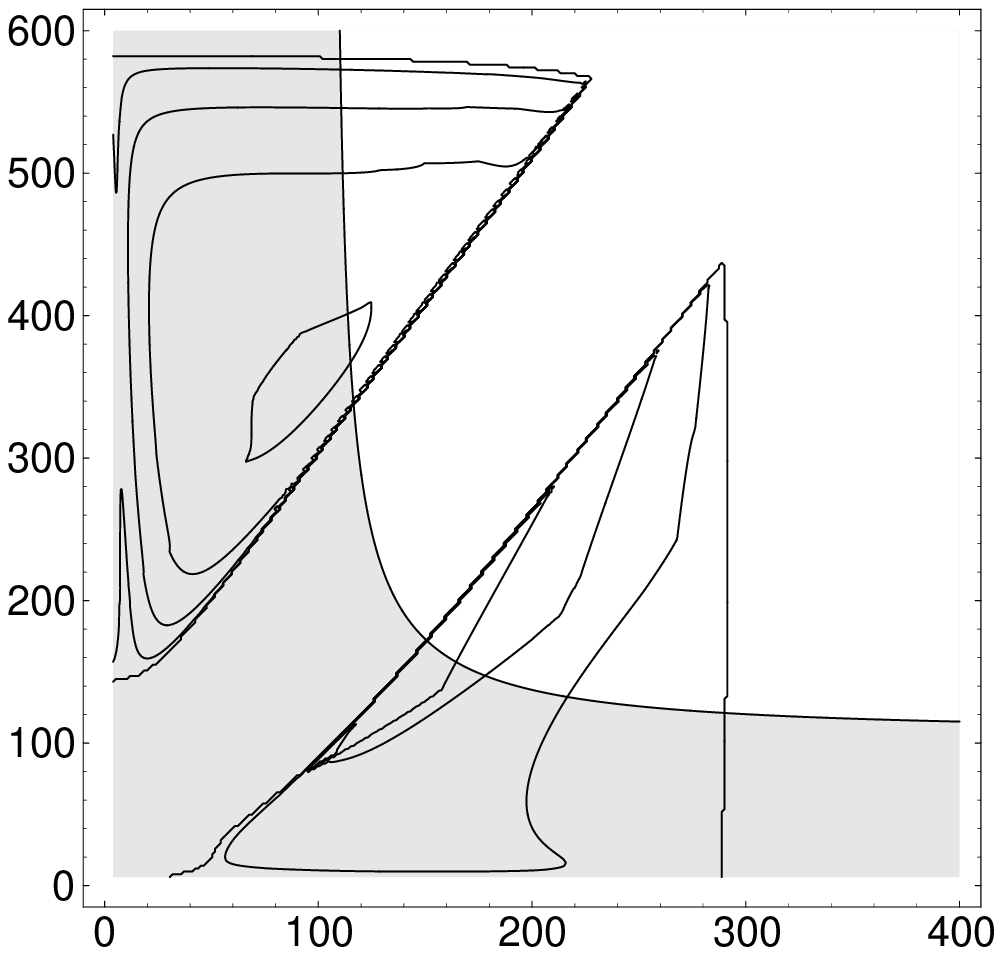}}
   \put(175,-8){$\mu$/GeV}
	\put(10,210){$M_2 /GeV$}
	\put(70,210){\fbox{BR($\tilde{\tau}_1 \;\to \;\tau  \;
			\tilde{\chi}^0_1 \;\ell  \;\bar \ell
$) in \% }}
\CArc(125,191)(5.5,0,380)
\Text(125,191)[c]{{\scriptsize B}}
\CArc(115,125)(6,0,380)
\Text(115,125)[c]{{\scriptsize A}}
\CArc(180,100)(6,0,380)
\Text(180,100)[c]{{\scriptsize B}}
		\put(65,120){{\scriptsize $0.6 $}}
		\put(87,159){{\scriptsize $0.3 $}}
		\put(60,171){{\scriptsize $0.1 $}}
		\put(40,180){{\scriptsize $0.01 $}}
		\put(111,76){{\scriptsize$0.3 $}}
		\put(132,87){{\scriptsize$0.1 $}}
		\put(132,63){{\scriptsize$0.01 $}}
		\put(34,-8){Fig.~\ref{stau3a}a}
\put(240,0){\includegraphics{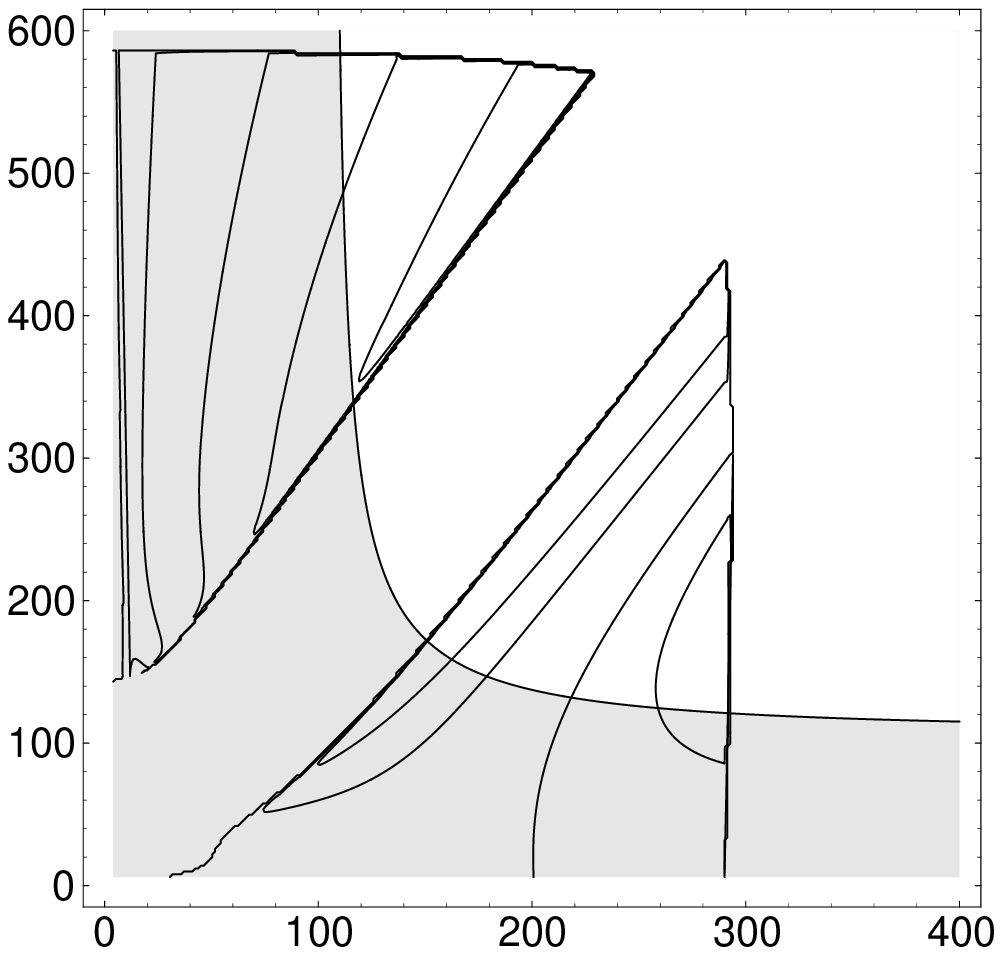}}
   \put(405,-8){$\mu$/GeV}
	\put(240,210){$M_2 /GeV$}
	\put(320,210){\fbox{${\mathcal A}_T^\ell$ in \% }}
\CArc(355,191)(5.5,0,380)
\Text(355,191)[c]{{\scriptsize B}}
\CArc(345,125)(6,0,380)
\Text(345,125)[c]{{\scriptsize A}}
\CArc(410,100)(6,0,380)
\Text(410,100)[c]{{\scriptsize B}}
	\put(275,160){{\scriptsize $0$}}
   \put(290,165){{\scriptsize $-1$}}
	\put(315,170){{\scriptsize $-2 $}}
	\put(340,175){{\scriptsize $-3$}}
		\put(337,67){{\scriptsize $1 $}}
		\put(323,40){{\scriptsize $ 0.5$}}
		\put(366,70){{\scriptsize $0 $}}
		\put(365,40){{\scriptsize$-0.2 $}}
		\put(260,-8){Fig.~\ref{stau3a}b }
 \end{picture}
\caption{Contour lines of the branching ratio 
for $\ti\tau_1\to \ti\chi^0_1\tau \ell\bar\ell$ 
and asymmetry $A_T^\ell$ defined in Eq.~\rf{eq:Toddasym} 
in the $M_2$-$\mu$ plane for $\varphi_{M_1}=\pi/2$ and $\varphi_{\mu}=0$,
$\tan\beta =10$, $A_{\tau}=1000$~GeV, 
$m_{\ti\tau_1}=300$~GeV, $ m_{\ti\tau_2}=800$~GeV for $M_{\ti E} < M_{\ti L}$.
The area A (B) is kinematically forbidden since  
$m_{\ti\chi_3^0}<m_{\ti\chi_1^0}+m_Z$
$(m_{\ti\tau_1}< m_{\ti\chi_3^0}+m_{\tau})$.  
The grey area is excluded since $m_{\chi^{\pm}_1}<104$~GeV.
\label{stau3a}}
\end{figure}

Next we discuss the decay chain 
$\ti \tau_1 \to \tau~\ti\chi^0_3 
\to \tau Z~\ti\chi^0_1 \to \tau \ti\chi^0_1\ell~\bar \ell$.
The two decays 
$\ti \tau_1 \to \tau~\ti\chi^0_2$ and
$\ti \tau_1 \to \tau~\ti\chi^0_3$ can be distinguished by 
measuring the $\tau$ energy in the $\T_1$ rest frame.
In Fig.~\ref{stau3a}a we show the contour
lines for the branching ratio BR$(\tau_1 \to \tau~\ti\chi^0_1~\ell~\bar \ell) 
={\rm BR}(\ti \tau_1 \to \tau~\ti\chi^0_3)\times
{\rm BR}(\ti\chi^0_3\to Z\ti\chi^0_1)\times
{\rm BR}(Z \to \ell~\bar \ell)$
in the $M_2$-$|\mu|$ plane
for $\varphi_{M_1}=\pi/2$ and $ \varphi_{\mu}=0$.
The area A (B) is kinematically forbidden since  
$m_{\ti\chi_3^0}<m_{\ti\chi_1^0}+m_Z$
$(m_{\ti\tau_1}< m_{\ti\chi_3^0}+m_{\tau})$.
The grey area is excluded since $m_{\chi^{\pm}_1}<104$~GeV.
We choose $M_{\ti E} < M_{\ti L}$ since the 
$\ti \tau_1$-$\tau$-$\ti\chi^0_3$ coupling $|a^{\ti\tau}_{13}|$ is larger, 
which implies a larger branching ratio 
BR$(\ti \tau_1 \to \tau~\ti\chi^0_3)$
than for $M_{\ti E} > M_{\ti L}$.
The total branching ratio is smaller than for the previous
decay chain since ${\rm BR}(\ti \tau_1 \to \tau~\ti\chi^0_3)<.75(0.05)$
in the upper (lower) part of Fig.~\ref{stau3a}a.

The corresponding asymmetry ${\mathcal A}_{\rm T}^\ell$ is shown in 
Fig.~\ref{stau3a}b. 
The asymmetry ${\mathcal A}_{\rm T}^\ell$
vanishes on contours where either 
$|a^{\ti\tau}_{13}|=|b^{\ti\tau}_{13}|$
or ${\rm Im}(O''^L_{13}{O''^R_{13}}^{\ast})=0$, 
see Eq.~\rf{eq:prop1}. On the one hand, along the contour line 0
in the lower part of  Fig.~\ref{stau3a}b we have 
$|a^{\ti\tau}_{13}|=|b^{\ti\tau}_{13}|$.
On the other hand, along the contour line 0 in the upper part
of Fig.~\ref{stau3a}b we have 
${\rm Im}(O''^L_{13}{O''^R_{13}}^{\ast})=0$.
Furthermore, between the upper and the lower part of Fig.~\ref{stau3a}b
(area A), there is a further sign change of 
${\rm Im}(O''^L_{13}{O''^R_{13}}^{\ast})$.
Concerning the first factor in  \rf{eq:prop1}, we remark that it
increases for  increasing $M_2$ and decreasing $|\mu|$.
This behaviour can be understood by observing
that for $|\mu|/M_2\to 0$ the gaugino component
of $\CH^0_3$ gets enhanced, resulting in 
$|b^{\ti\tau}_{13}|/|a^{\ti\tau}_{13}|\to 0$.

\begin{figure}[H]
\begin{picture}(120,220)(0,0)
\put(10,0){\includegraphics{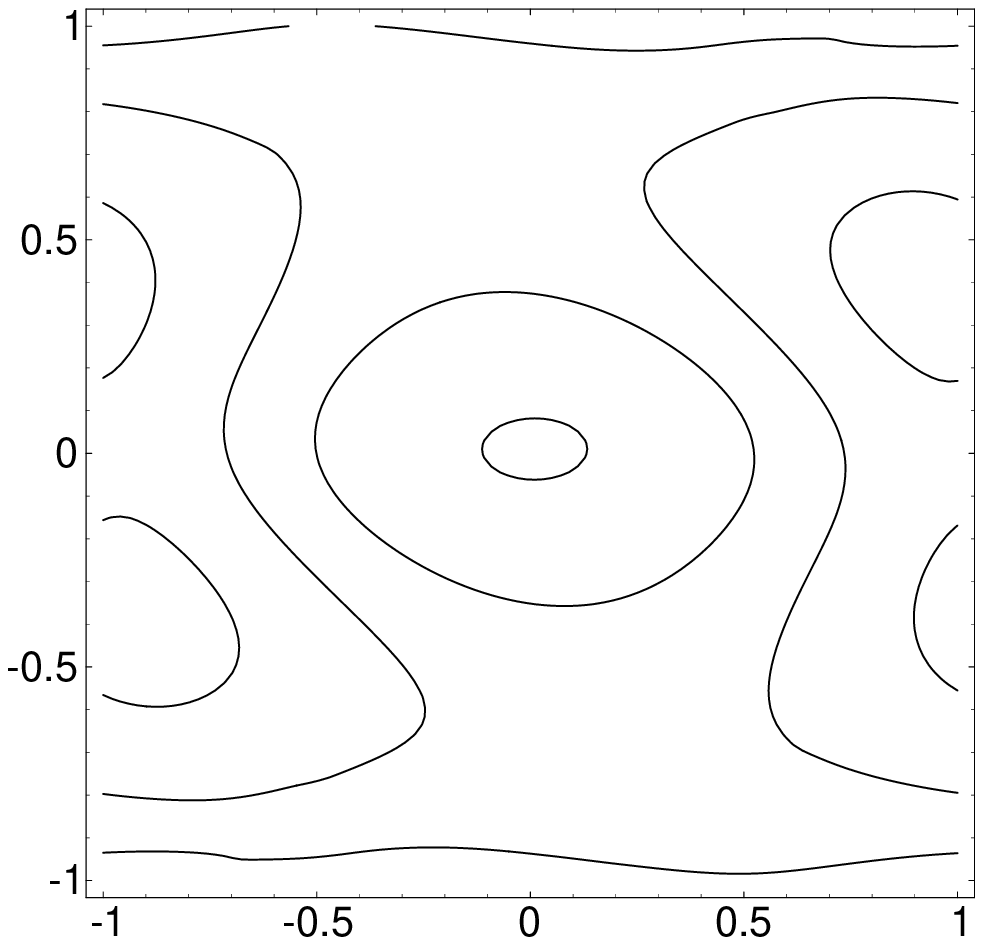}}
	\put(175,-8){$\varphi_{\mu}/\pi $}
	\put(10,210){$\varphi_{ M_1}/\pi$}
	\put(70,210){\fbox{BR($\tilde{\tau}_1 \;\to \;\tau  \;
				\tilde{\chi}^0_1 \;\ell  \; \bar \ell
		$) in \% }}
	\put(117,104){{\footnotesize $0.2 $}}
	\put(133,80){{\footnotesize $0.4 $}}
	\put(157,65){{\footnotesize $0.6 $}}
	\put(187,63){{\footnotesize $0.8 $}}
	\put(130,27){{\footnotesize $0.8 $}}
	\put(192,135){{\footnotesize $0.8 $}}
	\put(40,57){{\footnotesize $0.8 $}}
	\put(40,120){{\footnotesize $0.8 $}}
	\put(78,160){{\footnotesize $0.6 $}}
	\put(30,-8){Fig.~\ref{stau3b}a}
\put(240,0){\includegraphics{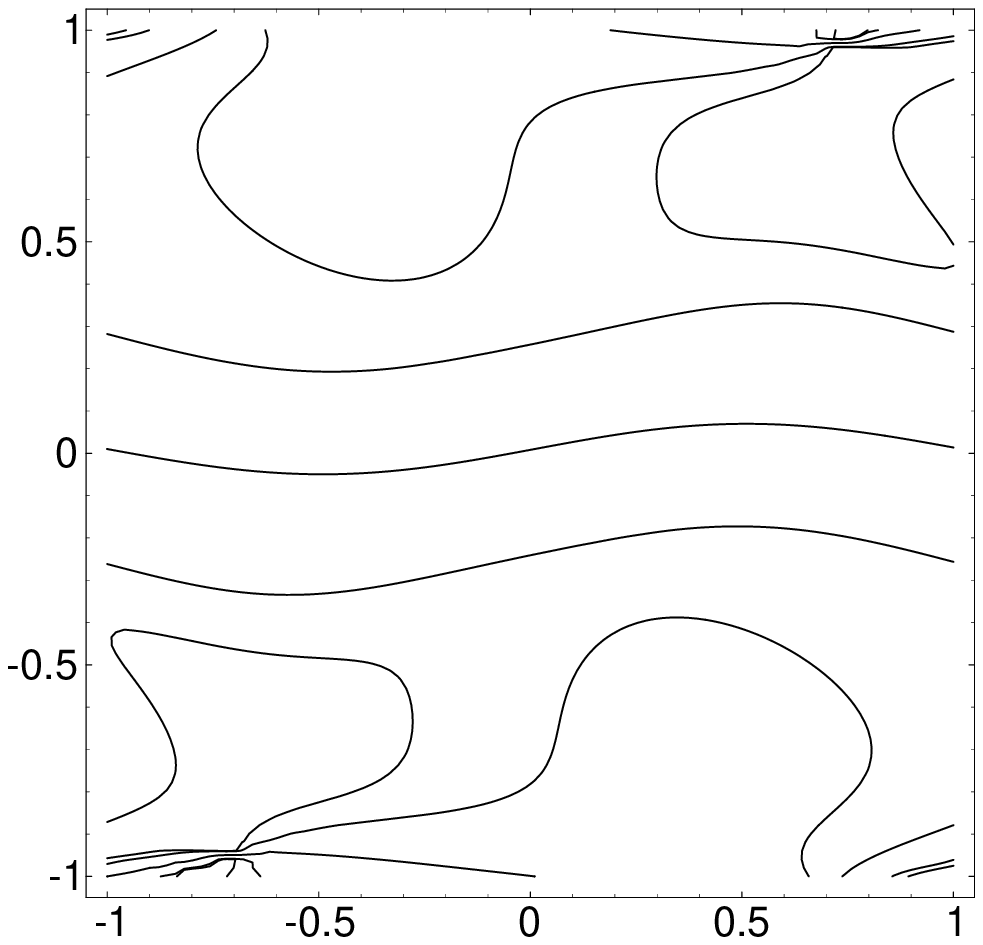}}
   \put(405,-8){$\varphi_{\mu}/\pi $}
	\put(240,210){$\varphi_{ M_1}/\pi$}
	\put(320,210){\fbox{${\mathcal A}_T^{\ell}$ in \% }}
        \put(294,127){{\footnotesize $-3 $}}
	\put(291,160){{\footnotesize $-3 $}}
	\put(400,153){{\footnotesize $-3.3 $}}
		\put(330,108){{\footnotesize $0 $}}
		\put(350,78){{\footnotesize $3 $}}
		\put(385,63){{\footnotesize $ 3 $}}
		\put(300,57){{\footnotesize $3.3 $}}
		\put(260,-8){Fig.~\ref{stau3b}b }
 \end{picture}
\caption{Contour lines of the branching ratio 
for $\ti\tau_1\to \ti\chi^0_1\tau \ell\bar\ell$ 
and asymmetry $A_T^\ell$ defined in Eq.~\rf{eq:Toddasym}
in the $\varphi_{ M_1}$-$\varphi_{\mu} $ plane for
$|\mu|=150$~GeV and $M_2=450$~GeV, taking
$\tan\beta =10$, $A_{\tau}=1000$~GeV, 
$m_{\ti\tau_1}=300$~GeV, $ m_{\ti\tau_2}=800$~GeV
for $M_{\ti E} < M_{\ti L}$.
\label{stau3b}}
\end{figure}

In Fig.~\ref{stau3b} we show 
the dependence of BR$(\tau_1 \to \tau~\chi^0_1~\ell~\bar \ell)$
and of ${\mathcal A}_{\rm T}^\ell$ on the phases 
$\varphi_{M_1}$ and $\varphi_{\mu}$,
fixing $|\mu|=150$~GeV and $M_2=450$~GeV.
For these parameters we
display in Table \ref{tab:chi3} the masses of $\ti\chi^0_i, i=1,\dots,4$ 
and the total widths $\Gamma_{\CH_3}$, $\Gamma_{\T_1}$ 
for various phase combinations.
Note that 
{\em maximal CP violating phases 
$\varphi_{\mu},\varphi_{M_1}=\pm\pi/2$
do not necessarily 
lead to the highest value of} ${\mathcal A}_{\rm T}^\ell$ due to the
complex interplay of the phases in 
${\rm Im}(O''^L_{13}{O''^R_{13}}^{\ast})$.
The value of ${\mathcal A}_{\rm T}^\ell$ depends stronger on $\varphi_{M_1}$,
which also determines the sign of ${\mathcal A}_{\rm T}^\ell$,
than on $\varphi_{\mu}$.
Comparing Fig.~\ref{stau2b}b and Fig.~\ref{stau3b}b, one can see 
that both figures have in common the strong $\varphi_{M_1}$
dependence, where in a good approximation
the sign of  ${\mathcal A}_{\rm T}^\ell$ is 
$sgn({\mathcal A}_{\rm T}^\ell)\approx sgn(\varphi_{M_1})$ 
in Fig.~\ref{stau2b}b and  
$sgn({\mathcal A}_{\rm T}^\ell)\approx -sgn(\varphi_{M_1})$ 
in Fig.~\ref{stau3b}b.
This difference can be traced back to the different
behaviour of ${\rm Im}(O''^L_{12}{O''^R_{12}}^{\ast})$ and
${\rm Im}(O''^L_{13}{O''^R_{13}}^{\ast})$.
Moreover, in  Fig.~\ref{stau3b}b two points of level crossing
appear at approximately 
$\varphi_{M_1}\approx \pm 0.95\pi$,
$\varphi_{\mu}\approx \pm 0.7\pi$.
\begin{table}[H]
\caption{
Masses of $\ti\chi^0_i, i=1,\dots,4$ 
and widths $\Gamma_{\CH_3}$, $\Gamma_{\T_1}$
for various phase combinations 
of $\varphi_{\mu}$ and $\varphi_{M_1}$, taking $|\mu|=150$~GeV and 
$M_2=450$~GeV,
$\tan\beta =10$, $A_{\tau}=1000$~GeV, 
$m_{\ti\tau_1}=300$~GeV, $ m_{\ti\tau_2}=800$~GeV for $M_{\ti E} < M_{\ti L}$.
}
\label{tab:chi3}
\begin{center}
\begin{tabular}{|c|c|c|c|c|} \hline
$\varphi_{\mu}$ & $\varphi_{M_1}$ &
$m_{\CH_1},m_{\CH_2},m_{\CH_3},m_{\CH_4}  ~[\rm GeV]$  &
$\Gamma_{\CH_3}~[\rm MeV]$ & $\Gamma_{\T_1}~[\rm MeV]$\\ \hline\hline
0 &            0    &$128,\;\;156,\;\;238,\;\;467$& 59.0 & 362 \\
0 & $\frac{\pi}{2}$ &$132,\;\;153,\;\;238,\;\;466$& 68.2 & 359 \\
0 & $ \pi   $       &$141,\;\;145,\;\;238,\;\;466$& 75.5 & 356 \\ \hline
$\frac{\pi}{2}$ &           0     & $131,\;\;158,\;\;237,\;\;466$& 41.5 &
356\\
$\frac{\pi}{2}$ & $\frac{\pi}{2}$ & $136,\;\;154,\;\;237,\;\;466$& 49.4 &
353  \\
$\frac{\pi}{2}$ & $\pi$           & $142,\;\;145,\;\;240,\;\;465$& 73.8 &
360  \\\hline
$\pi$ &       0         & $135,\;\;159,\;\;236,\;\;465$& 27.7  & 351 \\
$\pi$ & $\frac{\pi}{2}$ & $137,\;\;154,\;\;239,\;\;465$& 47.5  & 357 \\
$\pi$ & $\pi$           & $143,\;\;144,\;\;242,\;\;464$& 71.0  & 364 \\
\hline
 \end{tabular}
\end{center}
\end{table}

\vspace{2cm}

\section{Summary and Conclusion
   \label{summary}}

We have considered a T-odd correlation
and the corresponding asymmetry in the
sequential decay 
$\ti f \to f'~\ti\chi^0_j\to 
f'~\ti\chi^0_1~Z\to f'~\ti\chi^0_1~f \bar f$.
The analytical expressions have been given
in the density matrix formalism.
The contribution to the T-odd correlation
is induced by possible CP violating phases
in the neutralino sector.

In a numerical study of the decay
$\ti\tau_1\to\tau\chi^0_1~f \bar f$ we have shown that
the T-odd asymmetry considered can be of the order
of a few percent for leptonic final states.
The number of produced $\ti \tau$'s necessary to observe
${\mathcal A}_{\rm T}^\ell$ is at least of the
order $10^{5}$, which may be accessible at
future collider experiments.
For a semi-leptonic final state like
$\ti\chi^0_1~\tau~\bar b b$ the T-odd asymmetry
is larger by a factor 6.3.
If the T-odd asymmetry ${\mathcal A}_{\rm T}^b$
of such a semi-leptonic final state could be measured
with similar accuracy the 
number of produced $\ti \tau$'s necessary to observe
${\mathcal A}_{\rm T}^b$ is of the
order $10^{3}$.

\vskip10mm
\section*{Acknowledgements}

We thank S. Hesselbach and W. Majerotto for useful discussions.
This work is supported by the `Fonds zur
F\"orderung der wissenschaftlichen Forschung' (FWF) of Austria, projects
No. P13139-PHY and No. P16592-N02, by the European Community's 
Human Potential Programme
under contract HPRN--CT--2000--00149
and by Acciones Integradas Hispano--Austriaca.
T.K. acknowledges financial support from 
the European Commission Research Training Site contract HPMT-2000-00124.
O.K. was supported by the
\emph{Bayerische Julius-Maximilians Universit\"at W\"urzburg}.
This work was also supported by the Deutsche Forschungsgemeinschaft
(DFG) under contract Fr 1064/5-1.

\end{document}